\documentclass{article}
\usepackage{amsmath}
\usepackage{graphicx}
\usepackage{amsfonts}
\usepackage{amssymb}

\begin{document}

\title{\textbf{Symmetry, singularities and intregrability in complex dynamics VII:
Integrability Properties of FRW-Scalar Cosmologies}}
\author{Peter Leach\thanks{Permanent address: School of Mathematical and Statistical
Sciences, University of Natal, Durban 4041, Republic of South Africa}, Spiros
Cotsakis and John Miritzis\thanks{Department of Marine Sciences, University of
the Aegean, Sapfous 5, Mytilene 81 100{}, Greece}\\GEODESYC\\Department of Mathematics \\University of the Aegean\\Karlovassi 83 200, Greece}
\date{\today}
\maketitle
\begin{abstract}
\noindent This paper considers some physically interesting cosmological
dynamical systems in the FRW-scalarfield category which are examined for
integrability according to the criterion of Painlev\'e. In the literature
these systems have been examined from the point of view of dynamical systems
and the results from the two disparate methods of analysis are compared. This
allows some more general comments to be made on the use of the Painlev\'e
method in covariant systems.
\end{abstract}

\catcode`\@=11 \renewcommand{\theequation}{\thesection.\arabic{equation}}
\@addtoreset{equation}{section} \catcode`\@=10

%
%
%
%
%
%
%
%
%
%

%
%
%
%
%
%
%
%
%
%

%
%
%
%
%
%
%
%
%
%

%
%
%
%
%
%
%
%
%
%

%
%
%
%
%
%
%
%
%
%

%
%
%
%
%
%
%
%
%
%

%
%
%
%
%
%
%
%
%
%

%
%
%
%
%
%
%
%
%
%

%
%
%
%
%
%
%
%
%
%

%
%
%
%
%
%
%
%
%
%

%
%
%
%
%
%
%
%
%
%

%
%
%
%
%
%
%
%
%
%

%
%
%
%
%
%
%
%
%
%

%
%
%
%
%
%
%
%
%
%

%
%
%
%
%
%
%
%
%
%

%
%
%
%
%
%
%
%
%
%

\section{Introduction}

This is the seventh in a series of papers \cite{I,II,III,Miritzis 00,V,VI}
devoted to various aspects of the three mathematically disparate topics of
symmetry, singularities and integrability of dynamical systems. It is the
purpose of the series to cement the concepts into a greater obvious unity
whereby they are recognised as different aspects of the central properties of
dynamical systems which distinguish integrable systems from nonintegrable systems.

The plan of the present paper is as follows. In \S 2 we present a brief
resum\'{e} of the analysis of Painlev\'{e} with particular reference to some
of the subtleties required to deal with systems which do not fit into the neat
scheme of the standard ARS algorithm \cite{Ablowitz 78,Ablowitz 80a,Ablowitz
80b,Ramani 89}. In \S \S 3-6 we present the analysis of a number of models
chosen from some interesting areas of Cosmology which have been analysed from
the viewpoint of Dynamical Systems. In the approach used in Dynamical Systems
the original system of equations is replaced by an $n$-dimensional system of
first order equations, maybe subject to a constraint to preserve the actual
number of degrees of freedom. This is not always a suitable format for the
application of the Painlev\'{e} analysis and we find it necessary to rearrange
the systems into a more suitable format. The possession of the Painlev\'{e}
Property, i.e.,\textit{ } the existence of a solution analytic except at
moveable polelike singularities (branch point singularities in the case of the
`weak' property), is representation dependent, being preserved under a
homeographic transformation only, and consequently we must be careful in
comparing the results obtained using the Painlev\'{e} approach with those
obtained using the methods of Dynamical Systems. Finally, in \S \ref{concl},
we present some general observations of the results of our analysis paying
particular attention to the uses of Painlev\'{e} methods and analyses when
dealing with general relativistic systems.

\section{Methodology of singularity analysis}

For a detailed account of our approach to the Painlev\'e analysis the reader
is referred to our earlier paper \cite{Miritzis 00}. Here we present a summary
of that account so that the paper is self-contained for the reader not
concerned with the fine detail of the ideas behind the singularity analysis.

The basis of the singularity analysis of a system of differential equations is
found in the dominant behaviour of the system at a singularity and also the
next to leading order behaviour in the neighbourhood of the singularity
\cite{Feix 97,Leach 99}. When the analysis is concerned with the integrability
of a system, the conditions on the behaviour must be tightened to analytic
behaviour in the complex plane or at least substantial portions of it. (One
can treat with branch point singularities provided the branch cuts do not
become too concentrated in the region of the complex plane relevant to the
solution of the problem at hand.) This was the original approach by the
originators of this field -- Kowalevskaya \cite{Kowalevskaya} and Painlev\'{e}
\cite{Painleve 82, Painleve 93,Painleve 00}. The idea behind the analysis is
that a system of differential equations is integrable if there exists a set of
analytic functions which satisfy the system of differential equations. Since
an analytic function can be expressed in terms of Laurent series, the
demonstration that there exist Laurent expansions about movable singularities
containing the requisite number of arbitrary constants for the dependent
variables in the system of differential equations is taken to indicate that
the system is integrable. This is the basis of the ARS algorithm developed by
Ablowitz, Ramani and Segur \cite{Ablowitz 78,Ablowitz 80a,Ablowitz 80b} and
presented, for example, by Ramani \textit{et al }\ \cite{Ramani 89} and Tabor
\cite{Tabor 89}.

The essence of the ARS algorithm is that the solution of an $n$-dimensional
system of equations
\begin{equation}
\dot{x}_{i}=f_{i}(t,x), \label{2.1}%
\end{equation}
where the functions $f_{i}$ are rational in the dependent variables and
algebraic in the independent variable, can be written as either
\begin{equation}
x_{i}(t)=\sum_{j=0}^{\infty}a_{j}\tau^{-p_{i}+jq} \label{2.2}%
\end{equation}
in the case of a Right Painlev\'{e} Series or
\begin{equation}
x_{i}(t)=\sum_{j=0}^{\infty}a_{j}\tau^{-p_{i}-jq} \label{2.3}%
\end{equation}
in the case of a Left Painlev\'{e} Series \cite{Lemmer 93, Feix 97}, where
$\tau=t-t_{0}$ and $t_{0}$ is the location of the movable singularity, the
exponents $p_{i}$ are strictly positive integers (rational numbers in the case
of the so-called weak Painlev\'{e} test), the paramater $q$ is a positive
integer (respectively rational number) and in the coefficients there are $n-1$
arbitrary constants which, together with $t_{0}$, give the required number of
arbitrary constants for the general solution of the system (\ref{2.1}). In the
case of the Left Painlev\'{e} Series, which has the nature of an asymptotic
expansion in the region of $t$ infinite (\ref{2.3}) is usually written as
\begin{equation}
x_{i}(t)=\sum_{j=0}^{\infty}a_{j}t^{-p_{i}-jq} \label{2.4}%
\end{equation}
and the coefficients now contain $n$ arbitrary constants.

The ARS algorithm provides a mechanistic approach to the establishment of the
existence of (\ref{2.2}) and/or (\ref{2.4}), which is not always satisfactory
\cite{Hua 95}, but, when handled with sensitivity rather than automatically,
provides a useful tool in the analysis of differential equations. The first
step is to determine the leading order behaviour by means of the
\textit{Ansatz}
\begin{equation}
x_{i}=\alpha_{i}\tau^{-p_{i}},\label{2.4a}%
\end{equation}
which is substituted into the system (\ref{2.1}), and to assemble all possible
patterns of singular behaviour. For each of these possible patterns the
coefficients $\alpha_{i}$ are calculated. For each pattern the substitution
\begin{equation}
x_{i}=\alpha_{i}\tau^{-p_{i}}+\mu_{i}\tau^{r-p_{i}}\label{2.5}%
\end{equation}
is made to determine the `resonances' $r$ at which the arbitrary constants,
$\mu_{i}$, are introduced. These are determined from the leading order terms
of (\ref{2.1}) by a linearisation process.

For each pattern of singularity behaviour and its concomitant resonances and
arbitrary constants the substitution
\begin{equation}
x_{i}=\sum_{j=0}^{j=j_{max}}a_{i}\tau^{-p_{i}+jq},\label{2.6}%
\end{equation}
where $j_{max}q=r_{max}$ and $r_{max}$ is the largest resonance, is made into
the full system (\ref{2.1}) to ensure that there is consistency among the
resonances and with the nondominant terms of the original system. If this be
the case, that pattern of singular behaviour passes the Painlev\'{e} test and,
if this be the case for all possible patterns of singularity behaviour, the
system (\ref{2.1}) possesses the Painlev\'{e} Property which guarantees
integrability in most cases. There are exceptions to this general statement
and the interested reader is referred to Miritzis \textit{et al }%
\ \cite{Miritzis 00} and the references cited therein for a more detailed
discussion of the finer points of the application of the ARS algorithm to the
techniques of singularity analysis.

There are systems which cannot possess the Painlev\'e Property since they do
not have movable singularities in every variable. We speak not of those for
which one or more of the variables has a leading order behaviour in terms of a
positive rational exponent. This becomes singular after a sufficient number of
differentiations. Rather we refer to those variables which are constant at the
movable singularity, $t_{0}$, of other variables. (Positive integral behaviour
can be rectified by the simple, yet homeographic, means of taking the inverse
as a new variable.) Usually an analysis of these cases by means of the
\textit{Ansatz} of a series expansion -- the ARS algorithm in not applicable
in such a case and, indeed, is downright misleading -- produces series with
fewer that the required number of arbitrary constants. Naturally such series
cannot represent the general solution and give no indication of the
integrability of the system. However, subject to the convergence of the
series, they can indicate some sort of partial solution. For reasons explained
in Miritzis \textit{et al }\ \cite{Miritzis 00} we refer to these solutions as
`peculiar' solutions -- a literal translation from the Greek usage -- to avoid
some of the associations with other terms commonly used such as `singular' and
`particular'. A peculiar solution is one consisting of series which satisfy
the system (\ref{2.1}) containing fewer than the required number of arbitrary
constants. Finally we note that the considerations presented here in terms of
the systems of first order equations, (\ref{2.1}), apply \textit{mutatis
mutandis} to all systems of ordinary differential equations.

\section{Massive scalar field\label{sectmass}}

Consider the Friedmann equations for the evolution of the scale factor $a(t)$
of an open, flat or closed ($k=-1,0,+1$ respectively) universe in the case
where matter is represented by a self-interacting scalar field with potential
$V(\phi)$. They are
\begin{align}
\left(  \frac{\dot{a}}{a}\right)  ^{2}+\frac{k}{a^{2}}  &  =\frac{1}{3}\left(
\frac{1}{2}\dot{\phi}^{2}+V\left(  \phi\right)  \right) \nonumber\\
2\frac{\ddot{a}}{a}  &  =-\frac{1}{3}\left(  2\dot{\phi}^{2}-V\left(
\phi\right)  \right) \label{3.3}\\
\ddot{\phi}+3\frac{\dot{a}}{a}\dot{\phi}+V^{\prime}(\phi)  &  =0\nonumber
\end{align}
In this Section, we focus in the particular case that of the scalar potential,
$V(\phi)$, is quadratic of the form $\mbox{$\frac{1}{2}$}m^{2}\phi^{2}$ so
that the above equations become
\begin{align}
\left(  \frac{\dot{a}}{a}\right)  ^{2}+\frac{k}{a^{2}}  &  =\frac{1}{3}\left(
\frac{1}{2}\dot{\phi}^{2}+\frac{1}{2}m^{2}\phi^{2}\right) \nonumber\\
\frac{\ddot{a}}{a}  &  =-\frac{1}{6}\left(  2\dot{\phi}^{2}-\frac{1}{2}%
m^{2}\phi^{2}\right) \label{3.4}\\
\ddot{\phi}+3\frac{\dot{a}}{a}\dot{\phi}+m^{2}\phi &  =0.\nonumber
\end{align}
These three equations are not independent since, for example, the first and
third of (\ref{3.4}) lead to the second. (This is also the case in
(\ref{3.3}). The result is independent of the precise form of the scalar
field.) We make the choice (\ref{3.4}c) and so treat the system
\begin{align}
\left(  \frac{\dot{a}}{a}\right)  ^{2}+\frac{k}{a^{2}}  &  =\frac{1}{6}\left(
\dot{\phi}^{2}+m^{2}\phi^{2}\right) \nonumber\\
\ddot{\phi}+3\frac{\dot{a}}{a}\dot{\phi}+m^{2}\phi &  =0. \label{3.6}%
\end{align}
We can make some `cosmetic' improvements to (\ref{3.6}) by rescaling time,
$t$, scalar field, $\phi$ and scale factor (in the case $k\neq0$), $a$,
according to
\begin{equation}
\bar{t}=mt,\quad v^{2}=\mbox{$\frac{1}{6}$}\phi^{2}\quad\mbox{\rm and}\quad
u=\frac{ma^{2}}{k} \label{3.7}%
\end{equation}
to obtain the system
\begin{align}
\dot{u}^{2}  &  =u^{2}\left(  \dot{v}^{2}+v^{2}\right) \nonumber\\
u\ddot{v}+3\dot{u}\dot{v}+uv  &  =0 \label{3.8}%
\end{align}
in the case $k=0$ and the system
\begin{align}
\dot{u}^{2}+4u  &  =4u^{2}\left(  \dot{v}^{2}+v^{2}\right) \nonumber\\
2u\ddot{v}+3\dot{u}\dot{v}+2uv  &  =0 \label{3.9}%
\end{align}
in the case that $k\neq0$. In both instances the overdot now represents
differentiation with respect to the new time variable, $\bar{t}$. We observe
that the sign of $k$ is immaterial in the transformation (\ref{3.7}).

We commence the testing of (\ref{3.8}) for the possession of the Painlev\'{e}
Property with the standard substitution
\begin{equation}
a=\alpha\tau^{p}\qquad v=\beta\tau^{q},\label{3.10}%
\end{equation}
where $\tau=\bar{t}-\bar{t}_{0}$ and $\bar{t}_{0}$ represents the location of
the moveable singularity if $p$ is found to be negative, to determine the
leading order behaviour. We obtain the following exponents for the several
terms in (\ref{3.8}a) and (\ref{3.8}b) respectively
\begin{equation}%
\begin{array}
[c]{ccccc}%
2p-2 & \quad & 2p+2q-2 & \quad & 2p+2q\\
p+q-2 & p+q-2 & p+q. &  &
\end{array}
\label{3.11}%
\end{equation}
From (\ref{3.11}a) we see that $p$ is arbitrary and that balancing between the
first and third terms is possible for $q=-1$. However, then the second term
has the lower exponent $2p-4$ and cannot balance if one is thinking in terms
of a Right Painlev\'{e} Series. However, there is no problem if one thinks in
terms of the existence of a Left Painlev\'{e} Series since now the second term
vanishes more rapidly than the two dominant terms at infinity as required.
Unfortunately the third term of (\ref{3.11}b) is incompatible with a Left
Painlev\'{e} Series and one must conclude that $q=0$ which thereby ends the
validity of the next to leading order analysis. The standard procedure
outlined in the ARS algorithm ceases to be applicable.

For (\ref{3.9}) we apply the same methodology as for (\ref{3.8}) and obtain
the same dismal results. There is no possibility of consistent singular behaviour.

In the cases of both (\ref{3.8}) and (\ref{3.9}) one can look for the
existence of a standard series solution in the absence of the Laurent series
implicit in the Painlev\'{e} analysis. Indeed we find
\begin{align}
a  &  =a_{0}+a_{1}\bar{t}+\frac{3a_{0}^{2}b_{0}^{2}-2a_{1}^{2}}{2a_{0}}\bar
{t}^{2}+\ldots\nonumber\\
v  &  =b_{0}+\frac{\sqrt{a_{1}^{2}-a_{0}^{2}b_{0}^{2}}}{a_{0}}\bar{t}-\frac
{1}{2}\left(  b_{0}+3\frac{a_{1}}{a_{0}}b_{1}\right)  \bar{t}^{2}+...
\label{3.12}%
\end{align}
in the case of (\ref{3.8}) and
\begin{align}
u  &  =a_{0}+a_{1}\bar{t}+\left(  3a_{0}b_{0}^{2}-2-\frac{a_{1}^{2}}{4a_{0}%
}\right)  \bar{t}^{2}+...\nonumber\\
v  &  =b_{0}+\frac{\sqrt{4a_{0}+a_{1}^{2}-4a_{0}^{2}b_{0}^{2}}}{2a_{0}}\bar
{t}-\frac{1}{4}\left(  2b_{0}+3\frac{a_{1}}{a_{0}}b_{1}\right)  \bar{t}%
^{2}+... \label{3.13}%
\end{align}
in the case of (\ref{3.9}). The coefficients $a_{0}$, $b_{0}$ and $a_{1}$ are
arbitrary in both cases. Coefficients of higher powers of $\bar{t}$ become
somewhat more complicated expressions.

We conclude that the model of the massive scalar field with a quadratic
potential described by (\ref{3.6}) is not integrable in the sense of
Painlev\'e no matter the curvature (the change of variables in (\ref{3.7}) can
at most have the effect of a change from possession of the Painlev\'e Property
to possession of the `weak' Painlev\'e Property). This is of course not
surprising given that very little is now about the evolution of this system in general.

However, making the usual slow-roll approximation, which basically amounts to
neglecting the $\ddot{\phi}$ and kinetic energy terms from the Friedman
equations, that is
\begin{equation}
\dot{\phi}\sim\frac{V^{\prime}}{\sqrt{V}}%
\end{equation}
implies that in the quadratic potential case we are considering, the exponents
may balance with $p$ arbitrary and $q=-1$. This again does not obviously
produce some great difference in the possession of the Painlev\'{e} property
but it does show that at least the ARS algorithm can be applied. So we see
that precisely this condition, which is conducive to inflationary solutions,
makes some of the difficulties disappear.

\section{Models with an exponential potential}

\subsection{The scalar-vacuum case}

The equations for the simplest case, that of an FRW model with a single scalar
field with an exponential potential $V=e^{-\lambda\phi}$ and no other matter
source, are \cite{Halliwell}:
\begin{align}
\ddot{a}  &  =-\dot{a}^{2}-2\dot{\phi}^{2}+\frac{1}{2}\lambda\dot{a}\dot{\phi
}+1\nonumber\\
\ddot{\phi}  &  =\frac{1}{2}\lambda\dot{\phi}^{2}-3\dot{a}\dot{\phi}+\frac
{1}{2}\lambda\label{7.1}%
\end{align}
subject to the constraint
\begin{equation}
\dot{a}^{2}=\dot{\phi}^{2}+1-ke^{-2a+\lambda\phi}, \label{7.2}%
\end{equation}
where, as usual, $k=0,\pm1$ depending upon whether the scalar curvature is
zero, positive or negative. As we noted in \S\ref{sectmass} in the instance of
equations (\ref{3.3}), (\ref{7.1}) and (\ref{7.2}) are not independent since
the differentiation of (\ref{7.2}) and one of (\ref{7.1}) leads to the other.
However, (\ref{7.1}) and one of (\ref{7.1}) and (\ref{7.2}) are not equivalent
systems since one is of the fourth order and the other of the third order.

\subsubsection{The scalar-vacuum case as a third order system}

For our third order system we take (\ref{7.1}b) and (\ref{7.2}), i.e.,\textit{
} we consider the system
\begin{align}
\ddot{\phi} &  =\frac{1}{2}\lambda\dot{\phi}^{2}-3\dot{a}\dot{\phi}+\frac
{1}{2}\lambda\nonumber\\
\dot{a}^{2} &  =\dot{\phi}^{2}+1-ke^{-2a+\lambda\phi}.\label{7.3}%
\end{align}
We transform the system (\ref{7.3}) into one suitable for application of the
singularity analysis by means of the nonhomeographic transformation
\begin{equation}
u=e^{-2a+\lambda\phi}\qquad v=e^{\lambda\phi}\label{7.4}%
\end{equation}
to obtain
\begin{align}
2u\ddot{v}-3\dot{u}\dot{v}-\lambda^{2}uv &  =0\label{7.5}\\
\dot{u}^{2}v^{2}-2u\dot{u}v\dot{v}+\left(  1-\frac{4}{\lambda^{2}}\right)
u^{2}\dot{v}^{2} &  =4u^{2}v^{2}-4ku^{3}v^{2}%
\end{align}

The dominant expressions in the leading order analysis are
\begin{equation}
2\alpha\beta q(q-1)\tau^{p+q-2}-3\alpha\beta pq\tau^{p+q-2} \label{7.6}%
\end{equation}
and
\begin{equation}
\alpha^{2}\beta^{2}p^{2}\tau^{2p+2q-2}-2\alpha^{2}\beta^{2}pq\tau
^{2p+2q-2}+\left(  1-\frac{4}{\lambda^{2}}\right)  \alpha^{2}\beta^{2}%
q^{2}\tau^{2p+2q-2}+4k\alpha^{3}\beta^{2}p^{2}\tau^{3p+2q} \label{7.7}%
\end{equation}
for (\ref{7.5}a) and (\ref{7.5}b) respectively after the usual substitutions
are made. From (\ref{7.7}) we obtain $p=-2,q$ arbitrary and, using this in
(\ref{7.6}), we find that $q=-2,0$.

Firstly we consider the pattern of singular behaviour $(-2,-2)$. From
(\ref{7.7}) we obtain $\alpha=4/k\lambda^{2}$ and $\beta$ remains arbitrary.
When we make the substitution
\begin{equation}
u=\alpha\tau^{-2}+\mu\tau^{r-2}\qquad v=\beta\tau^{-2}+\nu\tau^{r-2},
\label{7.8}%
\end{equation}
the condition that the required constants are arbitrary is that
\begin{equation}
\left|
\begin{array}
[c]{ccc}%
6r & \quad & 2r(r-2)\\
-\displaystyle{\frac{16}{\lambda^{2}}} &  & \displaystyle{\frac{16r}%
{\lambda^{2}}}%
\end{array}
\right|  =0 \label{7.9}%
\end{equation}
which has the solutions $r=-1,0$. The first is generic, the second reflects
the arbitrariness of $\beta$ and there is no third so that the system cannot
pass the Painlev\'{e} test with this pattern of singularity behaviour. The
solution is peculiar (\textit{cf}\ \cite{Miritzis 00}) with the first few
terms being given by
\begin{align}
u  &  =\frac{4}{k\lambda^{2}\tau^{2}}\left(  1+\frac{\left(  \lambda
\tau\right)  ^{2}}{12}+\frac{\left(  \lambda\tau\right)  ^{4}}{240}+...\right)
\nonumber\\
v  &  =\frac{b_{0}}{\tau^{2}}\left(  1+\frac{\left(  \lambda\tau\right)  ^{2}%
}{12}+\frac{\left(  \lambda\tau\right)  ^{4}}{240}+...\right)  \label{7.10}%
\end{align}
which is suggestive of the solution of (\ref{7.5}) when $v=\mbox{\rm const}u$
and the pole is fixed at $t=0$.

The singularity pattern $(-2,0)$ is without the ambit of the Painlev\'{e}
test. If we make the \textit{Ansatz}
\begin{equation}
u=\sum_{i=0}a_{i}\tau^{i-2}\qquad v=\sum_{i=0}b_{i}\tau^{i}, \label{7.11}%
\end{equation}
the requirement that the coefficients of the first two terms balance gives
\begin{equation}
a_{0}b_{1}=0\qquad a_{0}^{2}b_{0}^{2}+ka_{0}^{3}b_{0}^{2}=0. \label{7.12}%
\end{equation}
The choice $a_{0}=0,b_{0}\neq0$ leads to nonzero even coefficients for the
$b_{i}$ and for all $a_{i}=0$. The reverse choice gives the only nonzero
coefficient to be $a_{0}$. Clearly both are just nonsense solutions.

From the point of view of a third order system the scalar-vacuum case does not
pass the Painlev\'e test, does not possess the Painlev\'e Property and so is
not integrable in the sense of Painlev\'e. However, we do note that there does
exist a peculiar solution of the type mentioned by Ince \cite[p 355]{Ince} in
the context of an equation discussed by Chazy in 1912. Naturally this solution
would be acceptable were the system of the second order. This leads to the
approach discussed in the following subsection.

\subsubsection{The scalar-vacuum case as a second order system plus a first
order system}

An alternate approach to the treatment of systems (\ref{7.1}) and (\ref{7.2})
is to regard (\ref{7.1}) as the two dimensional system of first order
equations
\begin{align}
\dot{x}  &  =-x^{2}+\frac{1}{2}\lambda xy-2y^{2}+1\nonumber\\
\dot{y}  &  =-3xy+\frac{1}{2}\lambda y^{2}+\frac{1}{2}\lambda, \label{73.1}%
\end{align}
where $x=\dot{a}$ and $y=\dot{\phi}$. This system can be treated independently
of (\ref{7.2}) since the latter contains the two constants of integration
following from the integration of $x$ and $y$. (We note that this approach was
used by Halliwell \cite{Halliwell} to treat (\ref{73.1}) as a two-dimensional
dynamical system.)

The usual substitution for the leading order behaviour of (\ref{73.1}) gives
the following patterns of exponents
\begin{equation}%
\begin{array}
[c]{cccc}%
p-1 & 2p & p+q & 2q\\
q-1 &  & p+q & 2q.
\end{array}
\label{73.2}%
\end{equation}
The values $p=q=-1$ are consistent for both sets. This is the only possible
pattern since, although $p=-1,q>-1$ is fine for the first set, it fails for
the second since the left side of (\ref{73.2}a) becomes zero and so no
balancing can occur and the reverse happens with $p>-1,q=-1$. The coefficients
of the leading order terms satisfy
\begin{align}
-\alpha &  =-\alpha^{2}+\frac{1}{2}\lambda\alpha\beta-2\beta^{2}\nonumber\\
-\beta &  =-3\alpha\beta+\frac{1}{2}\lambda\beta^{2}\label{73.3}%
\end{align}
from the second of which it follows that
\begin{equation}
\alpha=\frac{1}{3}+\frac{1}{6}\lambda\beta\label{73.4}%
\end{equation}
and from the first of which with (\ref{73.4})
\begin{equation}
\left(  \lambda^{2}-36\right)  \beta^{2}+4\lambda\beta+4=0
\end{equation}
We recall that $\lambda>0$. Consequently the coefficients of the leading order
terms are given by
\begin{align}
\alpha=\frac{1}{6}    \beta=-\frac{1}{6}\quad\mbox{\rm
if}\quad\lambda=6\label{73.6}\\
\alpha=\mp\beta ,\quad  \beta=\frac{-2}{\lambda\pm6}\quad\mbox{\rm if}%
\quad0<\lambda\neq6.\label{73.7}%
\end{align}
This means that for general $\lambda$ there are two possible solutions to be
considered for consistency with the requirements of the Painlev\'{e} test.

The resonances are the eigenvalues of the equation
\begin{equation}
\left[
\begin{array}
[c]{cc}%
r-1+2\alpha-\frac{1}{2}\beta\lambda & -\frac{1}{2}\lambda\alpha+4\beta\\
3\beta &  r-1+3\alpha-\lambda\beta
\end{array}
\right]  \left[
\begin{array}
[c]{c}%
\mu\\
\nu
\end{array}
\right]  =0\label{73.8}%
\end{equation}
which follows from the substitution
\begin{equation}
x=\alpha\tau^{-1}+\mu\tau^{r-1}\quad\mbox{\rm and}\quad y=\beta\tau^{-1}%
+\nu\tau^{r-1}\label{73.9}%
\end{equation}
into the dominant terms of the system (\ref{73.1}). We obtain the generic
$r_{1}=-1$ no matter the value of $\lambda$ and $r_{2}=\mbox{$\frac{2}{3}$}$
in the case that (\ref{73.6}) applies and $r_{2}=\pm8/(\lambda\pm6)=\mp4\beta$
in the case that (\ref{73.7}) applies. Since the nondominant terms in
(\ref{73.1}) are constants, the expansion must be in a power of $\tau$ such
that it be commensurate with 1 and the second resonance, $r_{2}$. This
involves a constraint on the permissible values which $\lambda$ may take. It
is impossible for both values of $r_{2}$ to belong to the interval $(0,1)$ and
so in this range the system (\ref{73.1}) cannot pass the Painlev\'{e} test.
However, the requirements of commensurability and two positive values of
$r_{2}$ can be satisfied for some $\lambda\in(0,6)$ and the potential of both
branches to pass the test is present. The possession of the Painlev\'{e}
Property by (\ref{73.1}) is not possible in the case $r_{2}=1$, i.e.,\textit{
} $\lambda=2$ since at that value the arbitrary constant that is introduced
has zero coefficient and cannot be used to compensate for the presence of the
nondominant constant term in (\ref{73.1}). The constant terms deny the
possibility of a Left Painlev\'{e} series and are responsible for the
restriction on the interval of $\lambda$ for which interval one can hope to
obtain Right Painlev\'{e} series for both solutions.

Note that possession of the Painlev\'{e} Property by (\ref{73.1}) means that
the original second order pair (\ref{7.1}) does not possess it since both $a$
and $\phi$ will contain a logarithmic singularity due to the quadrature of the
leading order terms. Nevertheless a solution of lessened analytic quality will
exist and one must now consider the effect of the constraint (\ref{7.2}).
After a little manipulation the combination $x\mbox{\rm(\ref{73.1}a)}%
-y\mbox{\rm
(\ref{73.1}b)}$ leads to
\begin{equation}
x\dot{x}-y\dot{y}=\frac{1}{2}\left(  x^{2}-y^{2}-1\right)  \left(  -2x+\lambda
y\right)  \label{73.10}%
\end{equation}
which is immediately integrable to
\begin{equation}
x^{2}-y^{2}-1=cst\exp(-2a+\lambda\phi). \label{73.11}%
\end{equation}
Consequently there is a further constraint on the coefficients of the
expansions for $x(\tau)$ and $y(\tau)$ determined by the constant of
integration in (\ref{73.11}). This constant is the value of $k$, $=0,\pm1$,
and this will fix the single arbitrary constant found in the series (apart
from the location of the singularity).

For example, in the case that $\lambda=10$ so that $r_{2}=1/2$, the expansions
are
\begin{equation}
x\left(  t\right)  =\sum_{n=0}^{\infty}a_{n}\tau^{n/2-1},\qquad y\left(
t\right)  =\sum_{n=0}^{\infty}b_{n}\tau^{n/2-1}. \label{73.12}%
\end{equation}

The result is surprisingly nice as the first few coefficients are
\begin{align}
a_{0}=1/8,\;\;\;\;\;\;\;\;\;\;\;\;\;\;\;  &  b_{0}=-1/8\nonumber\\
a_{1}=3b_{1},\;\;\;\;\;\;\;\;\;\;\;\;\;\;\;  &  b_{1}=b_{1}\nonumber\\
a_{2}=2b_{1}^{2},\;\;\;\;\;\;\;\;\;\;\;\;\;\;\;  &  b_{2}=-2b_{1}%
^{2}\nonumber\\
a_{3}=-24b_{1}^{3},\;\;\;\;\;\;\;\;\;\;\;\;  &  b_{3}=-8b_{1}^{3}\nonumber\\
a_{4}=\frac{1}{6}\left(  11-192b_{1}^{4}\right)  ,\quad &  b_{4}=\frac{1}%
{6}\left(  13+192b_{1}^{4}\right)  \label{73.13}%
\end{align}
and we obtain
\begin{align}
x\left(  t\right)   &  =\frac{1}{8}\left(  t-t_{0}\right)  ^{-1}+3b_{1}\left(
t-t_{0}\right)  ^{-1/2}+2b_{1}^{2}-24b_{1}^{3}\left(  t-t_{0}\right)
^{1/2}+O\left(  t-t_{0}\right) \nonumber\\
y\left(  t\right)   &  =-\frac{1}{8}\left(  t-t_{0}\right)  ^{-1}+b_{1}\left(
t-t_{0}\right)  ^{-1/2}-2b_{1}^{2}-8b_{1}^{3}\left(  t-t_{0}\right)
^{1/2}+O\left(  t-t_{0}\right)  .\nonumber\\
&
\end{align}
Substitution in the last constraint equation seems to be a formidable task.
The left hand side is
\begin{align}
LHS  &  =b_{1}\left(  t-t_{0}\right)  ^{-3/2}+8b_{1}^{2}\left(  t-t_{0}%
\right)  ^{-1}+8b_{1}^{3}\left(  t-t_{0}\right)  ^{-1/2}-1-128b_{1}%
^{4}\nonumber\\
&  -128b_{1}^{5}\left(  t-t_{0}\right)  ^{1/2}+512b_{1}^{6}\left(
t-t_{0}\right)  +O\left(  t-t_{0}\right)  ^{3/2} \label{73.14}%
\end{align}
and the right hand side, up to the constant of integration,
\begin{align}
RHS  &  =\left(  t-t_{0}\right)  ^{-3/2}+8b_{1}\left(  t-t_{0}\right)
^{-1}+8b_{1}^{2}\left(  t-t_{0}\right)  ^{-1/2}-128b_{1}^{3}\nonumber\\
&  -480b_{1}^{4}\left(  t-t_{0}\right)  ^{1/2}+\mbox{$\frac{1792}{5}$}%
b_{1}^{5}\left(  t-t_{0}\right)  +O\left(  t-t_{0}\right)  ^{3/2}.
\label{73.15}%
\end{align}

Consequently and therefore we can only conclude that, although there is
agreement in the two series for the initial terms, this agreement fails for
higher terms in the Laurent expansions. In the neighbourhood of the
singularity the two would appear to be almost equal and in a numerical
integration one could be tempted to ascribe the discrepancies to numerical
error and to input integrability in the sense of Painlev\'{e} to a systems
which is not integrable in that sense. (One recalls a similar suggestion of
integrability in the surface of section plots for the H\'{e}non-Heiles problem
at low energies, the equivalent to being close to the singularity in the
present instance.)

\subsection{The scalar-fluid case}

This is the case where a general (flat or curved) FRW model has a scalar field
(with an exponential potential) and a separately conserved perfect fluid. The
equations for the flat case are \cite{gr-qc/9711068} [Eqns (6)-(7)]
\begin{align}
x^{\prime} &  =-\frac{3}{2}(2-\gamma)x+\lambda\sqrt{\frac{3}{2}}%
y^{2}-3(2-\gamma)x^{3}-3\gamma xy^{2}\label{8.1}\\
y^{\prime} &  =-\frac{3}{2}y-\lambda\sqrt{\frac{3}{2}}xy-3(2-\gamma
)x^{2}y-3\gamma y^{3}\label{8.2}%
\end{align}
which are derived from the system
\begin{align}
\dot{h} &  =-\frac{1}{2}\kappa^{2}\left(  \rho_{\gamma}+p_{\gamma}+\dot{\phi
}^{2}\right)  \nonumber\\
\dot{p}_{\gamma} &  =-3H\left(  \rho_{\gamma}+p_{\gamma}\right)  \nonumber\\
\ddot{\phi} &  =-3H\dot{\phi}-V^{\prime}(\phi)\label{8.3}%
\end{align}
subject to
\begin{align}
p_{\gamma} &  =(\gamma-1)\rho_{\gamma}\\
V &  =V_{0}e^{-\lambda\kappa\phi}\\
H^{2} &  =\frac{1}{3}\kappa^{2}\left(  \rho_{\gamma}+\frac{1}{2}\dot{\phi}%
^{2}+V\right)  \label{8.4}%
\end{align}
which represent the assumed equation of state, the assumed form of the
potential and the constraint following from the Einstein Field Equations, by
the definitions
\begin{equation}
x=\frac{\dot{\phi}\kappa}{\sqrt{6}H},\quad y=\frac{\kappa\sqrt{V}}{\sqrt{3}%
H},\quad\tau=\ln a\label{8.5}%
\end{equation}
and the prime denotes differentiation with respect to new time, $\tau$. In
terms of the solutions of the system (\ref{8.1}) and (\ref{8.2}) the
constraint equation (\ref{8.4}) now becomes the definition of the variable
$\kappa\sqrt{\rho_{\gamma}}/\sqrt{3}H$ and so the treatment of the system
(\ref{8.1}) and (\ref{8.2}) can be undertaken without the necessity to
consider any constraint.

We substitute
\begin{equation}
x=\alpha\tau^{p}\qquad y=\beta\tau^{q}\label{8.6}%
\end{equation}
to obtain the set of exponents
\begin{equation}%
\begin{array}
[c]{ccccc}%
p-1 & p & 2q & 3p & p+2q\\
q-1 & q & p+q & 2p+q & 3q
\end{array}
\label{8.7}%
\end{equation}
from which it is evident that there are two possible patterns of leading order
behavior, \textit{viz } $p=q=-\mbox{$\frac{1}{2}$}$ involving the left side
and final two terms of both equations and $p=0$, $q=-\mbox{$\frac{1}{2}$}$
involving the second and fourth terms of the right side of (\ref{8.1}) and the
left side and final term of the right side of (\ref{8.2}). The second pattern
does not fit into the scheme of the Painlev\'{e} test and so the possession of
the Painlev\'{e} Property depends only upon the first pattern. We note that
the third order terms in both equations can also balance since they share the
symmetry $x\partial_{x}+y\partial_{y}$ \cite{Leach 99}, but this leads to some
constant solutions which are not comparable with the remaining terms in the
two equations.

Case $p=q=-\mbox{$\frac{1}{2}$}$:

From the dominant terms of both equations the coefficients are required to
satisfy the single equation
\begin{equation}
\gamma\beta^{2}=(2-\alpha)^{2}-\frac{1}{3}, \label{8.8}%
\end{equation}
which is very encouraging for the passing of the Painlev\'{e} test. To
determine the resonances we substitute
\begin{equation}
x=\alpha\tau^{-\mbox{$\frac{1}{2}$}}+\mu\tau^{r-\mbox{$\frac{1}{2}$}}\qquad
y=\beta\tau^{-\mbox{$\frac{1}{2}$}}+\nu\tau^{r-\mbox{$\frac{1}{2}$}}
\label{8.9}%
\end{equation}
into the dominant terms to obtain the two dimensional system
\begin{equation}
\left[
\begin{array}
[c]{ccc}%
r+1-3\gamma\beta^{2}, & \quad & 3\alpha\beta\gamma\\
-3(2-\gamma), &  & r+3\gamma\beta^{2}%
\end{array}
\right]  \left[
\begin{array}
[c]{c}%
\mu\\
\nu
\end{array}
\right]  =0 \label{8.10}%
\end{equation}
in which (\ref{8.8}) has been used to make some simplification.

The system (\ref{8.10}) has a nontrivial solution if $r=-1,0$ which is
consistent with (\ref{8.8}) and means that the system (\ref{8.1}) and
(\ref{8.2}) possesses the (weak) Painlev\'{e} Property since the nondominant
terms do not enter until higher powers.

For the other pattern of behaviour we substitute
\begin{equation}
x = \sum_{i=0}^{\infty}\tau^{\mbox{$\frac{1}{2}$} i}\quad\mbox{\rm and}\quad y
= \sum_{i=0}^{\infty} \tau^{\mbox{$\frac{1}{2}$} i-\mbox{$\frac{1}{2}$}}.
\label{8.11}%
\end{equation}
We find that the odd coefficients are all identically zero and that the only
arbitrary constant is the location of the movable singularity (in $y$),
$t_{0}$. Consequently we have a peculiar solution.

In terms of the definitions of the new variables in terms of the original
variables the constraint equation, (\ref{8.4}), gives $\kappa\sqrt
{\rho_{\gamma}}/\sqrt{3}H$ as a function with a square root singularity at
$t_{0} $ analytic, away from the single branch cut, in the variable $\tau$.
Unfortunately the definition of new time makes all functions nonanalytic in
terms of the original time variable. However, there is a sense of
integrability provided one keeps to the new time. It is not possible to make
even this remark about the original dependent variables since it is the three
combinations of them which have been shown to have the moderately good behaviour.

The equations for the case where a general curved FRW model has a scalar field
with an exponential potential and a separately conserved perfect fluid are
given by van den Hoogen \textit{et al }\ \cite{Hoogen} [Eqns (1.10-1.12)].
From the fourth order system
\begin{align}
\dot{H}  &  =-\frac{1}{2}\left(  \gamma\rho_{\gamma}+\dot{\phi}^{2}\right)
-K\nonumber\\
\dot{\rho}_{\gamma}  &  =-3\gamma H\rho_{\gamma}\nonumber\\
\ddot{\phi}  &  =-3H\dot{\phi}+\kappa V \label{8.12}%
\end{align}
subject to the constraint
\begin{equation}
H^{2}=\frac{1}{3}\left(  \rho_{\gamma}+\frac{1}{2}\dot{\phi}^{2}+V\right)  +K
\label{8.12a}%
\end{equation}
we obtain the third order system
\begin{align}
x^{\prime}  &  =-2x+Ay^{2}+Bx\Omega+2x^{3}-xy^{2}\label{8.13}\\
y^{\prime}  &  =y-Axy+By\Omega+2x^{2}y-y^{3}\label{8.14}\\
\Omega^{\prime}  &  =-2B\Omega+2B\Omega^{2}+4x^{2}\Omega-2y^{2}\Omega,
\label{8.15}%
\end{align}
in which we have written $A=\kappa\sqrt{\frac{3}{2}}$ and $B=\frac{3}{2}%
\gamma-1$to make the expressions look slightly simpler.

Under the usual substitution for the leading order behaviour we find the
following pattern of exponents
\begin{equation}%
\begin{array}
[c]{cccccc}%
p-1 & p & 2q & p+r & 3p & p+2q\\
q-1 & q & p+q & q+r & 2p+q & 3q\\
r-1 & r &  & 2r & 2p+r & 2q+r.
\end{array}
\label{8.16}%
\end{equation}
It is evident that the general singularity pattern is $p=q=-\mbox{$\frac{1}%
{2}$},r=-1$. There are also other patterns of a more specialised nature. We
list them in Table I.

\begin{center}%
\begin{equation}%
\begin{array}
[c]{|c|c|c|l|}\hline
p & q & r & \mbox{\rm Observation}\\\hline
-\mbox{$\frac{1}{2}$} & >-\mbox{$\frac{1}{2}$} & >-1 & \mbox{\rm not P}\\
>-\mbox{$\frac{1}{2}$} & -\mbox{$\frac{1}{2}$} & >-1 & \mbox{\rm not P}\\
>-\mbox{$\frac{1}{2}$} & >-\mbox{$\frac{1}{2}$} & -1 & \mbox{\rm not P}\\
-\mbox{$\frac{1}{2}$} & -\mbox{$\frac{1}{2}$} & >-1 & \text{P if }r=-\frac
{1}{2}\\
-\mbox{$\frac{1}{2}$} & >-\mbox{$\frac{1}{2}$} & -1 & \mbox{\rm not P}\\
>-\mbox{$\frac{1}{2}$} & -\mbox{$\frac{1}{2}$} & -1 & \mbox{\rm not P.}%
\\\hline
\end{array}
\nonumber
\end{equation}
\begin{minipage}{100mm} Table I: Possible sundominant patterns of semisingular
behaviour  for the system \ref{8.13} - \ref{8.15}.\end{minipage}
\end{center}

Only the fourth case is a candidate for the Painlev\'{e} test as the other
possibilities do not conform to the singularity requirement of the function
and/or its derivatives.

In the general case the coefficients of the leading order terms satisfy the
single constraint
\begin{equation}
4\alpha^{2}-2\beta^{2}+2B\gamma+1=0 \label{8.17}%
\end{equation}
so that one would expect to find the resonances $r=-1,0(2)$. With the use of
(\ref{8.17}) the characteristic equation for the resonances is
\begin{align}
\left|
\begin{array}
[c]{ccc}%
r-4\alpha^{2} & 2\alpha\beta & -B\alpha\\
-4\alpha\beta &  r+2\beta^{2} & -B\beta\\
-8\alpha\gamma & 4\beta\gamma &  r-2B\gamma
\end{array}
\right|   &  =0\nonumber\\
& \nonumber\\
\Leftrightarrow r^{3}+r^{2}  &  =0
\end{align}
so that $r=-1,0(2)$ as anticipated from (\ref{8.17}) so that there is
consistency. There are three arbitrary constants and the nondominant terms can
cause no theoretical difficulties. For this pattern of singular behaviour
(\ref{8.13}) - (\ref{8.15}) passes the Painlev\'e test.

The other candidate for the Painlev\'{e} test, case four of Table I, with
$p=q=r=-\mbox{$\frac{1}{2}$}$ produces the system
\begin{align}
-\frac{1}{2}\alpha &  =2\alpha^{3}-\alpha\beta^{2}\nonumber\\
-\frac{1}{2}\beta &  =2\alpha^{2}\beta-\beta^{3}\nonumber\\
-\frac{1}{2}\gamma &  =4\alpha^{2}\gamma-2\beta^{2}\gamma\label{8.18}%
\end{align}
for the coefficients of the leading order terms. There is a contradiction
between the first two equations of (\ref{8.18}) on the one hand and the third
equation on the other. Consequently this is not an admissible case for the
system (\ref{8.13}) - (\ref{8.15}). The one admissible singularity pattern has
passed the Painlev\'{e} test and so we conclude that the system (\ref{8.13}) -
(\ref{8.15}) possesses the (weak) Painlev\'{e} Property. The constraint
equation, (\ref{8.12a}), for the original fourth order system. (\ref{8.12}),
defines the fourth variable. The same comments as for the system (\ref{8.3})
apply to this case also.

\section{The general flat case}

We use the notation of Foster \cite{Foster}. He presents the dynamical system
\begin{align}
\frac{d\phi}{dt}  &  =\dot{\phi}\nonumber\\
\frac{d\dot{\phi}}{dt}  &  =-K\dot{\phi}-V^{\prime}(\phi)\nonumber\\
\frac{dK}{dt}  &  =-\frac{3}{2}\dot{\phi}^{2} \label{9.1}%
\end{align}
subject to the constraint
\begin{equation}
K^{2}=3V(\phi)+\frac{3}{2}\dot{\phi}^{2}. \label{9.2}%
\end{equation}
Foster introduces the new variables
\begin{equation}
x=\frac{1}{K},\quad y=\sqrt{\frac{3}{2}}\frac{\dot{\phi}}{K}\;\;\text{and
\ \ }\tau=\log v\left(  t\right)  +\tau_{0} \label{9.3}%
\end{equation}
where $K$ is the trace of the extrinsic curvature and $v(t)$
represents volume, $v\sim a^3$. Then the system becomes
\begin{align}
\frac{dx}{d\tau}  &  =y^{2}x\nonumber\\
\frac{dy}{d\tau}  &  =-y-\sqrt{\frac{3}{2}}x^{2}V^{\prime}\left(  \phi\right)
+y^{3}\nonumber\\
\frac{d\phi}{d\tau}  &  =\sqrt{\frac{2}{3}}y \label{9.4}%
\end{align}
and the constraint is now
\begin{equation}
y^{2}+3x^{2}V(\phi)=1. \label{9.5}%
\end{equation}

In order to make some reasonable progress with the Painlev\'{e} analysis of
this system we require some explicit functional form for the potential,
$V(\phi)$. (We note that in Foster's Dynamical Systems approach \cite{Foster}
such a constraint was not necessary. All that was required were some general
properties of the potential. The approaches via Dynamical Systems and
Singularity Analysis are complementary.) We make the standard \textit{Ansatz}
for a confining potential (\textit{cf}\ the transition from (\ref{3.3}) to
(\ref{3.4})), \textit{viz }
\begin{equation}
V(\phi)=\mbox{$\frac{1}{2}$}m^{2}\phi^{2} \label{9.6}%
\end{equation}
so that the system (\ref{9.5}) is now
\begin{align}
\frac{dx}{d\tau}  &  =y^{2}x\nonumber\\
\frac{dy}{d\tau}  &  =-y-\sqrt{\frac{3}{2}}x^{2}m^{2}\phi+y^{3}\nonumber\\
\frac{d\phi}{d\tau}  &  =\sqrt{\frac{2}{3}}y \label{9.7}%
\end{align}
and the constraint is now
\begin{equation}
y^{2}+\frac{3}{2}x^{2}m^{2}\phi^{2}=1. \label{9.8}%
\end{equation}

We make the usual substitution for the leading order behaviour to obtain
\begin{align}
\alpha p\tau^{p-1}  &  =\alpha\beta^{2}\tau^{p+2q}\label{9.9.1}\\
\beta q\tau^{q-1}  &  =-\beta\tau^{q}+\beta^{3}\tau^{3q}-\sqrt{\frac{3}{2}%
}\alpha^{2}\gamma^{2}m^{2}\tau^{2p+r}\label{9.9.2}\\
\gamma r\tau^{r-1}  &  =\sqrt{\frac{2}{3}}\beta\tau^{q} \label{9.9.3}%
\end{align}

From (\ref{9.9.1}) we have: $q=-\mbox{$\frac{1}{2}$}$, $\alpha$ is
arbitrary, $p$ arbitrary and $\beta^{2}=p$. With this
(\ref{9.9.3}) gives $r=\mbox{$\frac{1}{2}$}$ and
$\gamma=2\sqrt{2/3}\beta$. In (\ref{9.9.2}) the possibility that
$p<-1$ cannot occur as there would only be one term and so there
can be no balance. If $p>-1$, the two dominant terms require that
$\beta^{2}=-\mbox{$\frac{1}{2}$}$ and from above this means
$p=-\mbox{$\frac{1}{2}$}$, but $\alpha$ remains arbitrary. If
$p=-1$, $\alpha^{2}=-1/4m^{2}$.

The system (\ref{9.7}) permits the two patterns of dominant behaviour,
\textit{viz } \newline$(-\mbox{$\frac{1}{2}$},-\mbox{$\frac{1}{2}$},\mbox{$\frac{1}%
{2}$})$ and $(-1,-\mbox{$\frac{1}{2}$},\mbox{$\frac{1}{2}$})$. For
the first pattern of dominant behaviour the resonances are found
to be $r=-1,-\mbox{$\frac{1}{2}$},0$. The first is the generic
value. The third reflects the arbitrary value of $\alpha$. The
second demonstrates that this pattern cannot pass the Painlev\'{e}
test since it requires a Left Painlev\'{e} Series and this is
inconsistent with the nondominant terms in the original system
(\ref{9.7}) which require a Right Painlev\'{e} Series. The series
expansion for this pattern of dominant behaviour contains only two
arbitrary constants and consists of even terms only. There are two
expansions depending upon whether a positive or negative root is
taken. The first few terms of the series based on the positive
root are given by
\begin{align}
x &  =a_{0}\tau^{-\mbox{$\frac{1}{2}$}}+\frac{1}{2}a_{0}\left(  1+2m^{2}%
a_{0}^{2}\right)  \tau^{-\mbox{$\frac{1}{2}$}}\nonumber\\
y &  =\frac{i}{\sqrt{2}}\tau^{-\mbox{$\frac{1}{2}$}}-\frac{i\sqrt{2}}%
{4}\left(  1+2m^{2}a_{0}^{2}\right)  \tau^{-\mbox{$\frac{1}{2}$}}\nonumber\\
\phi &  =\frac{2i}{\sqrt{3}}\tau^{\mbox{$\frac{1}{2}$}}-\frac{i\sqrt{3}}%
{9}\left(  1+2m^{2}a_{0}^{2}\right)  \tau^{-\mbox{$\frac{3}{2}$}}\label{9.10}%
\end{align}

For the second pattern of dominant behaviour we find that $r=-1(3)$ which
means that there is only one arbitrary constant, the location of the moveable
singularity. The series expansion consists of even terms only and for the
positive root the first few terms are
\begin{align}
x  &  =\frac{i}{2m}\tau^{-1}+\frac{3i}{16m}+\frac{155-243i}{9216m}\tau
+\ldots\nonumber\\
y  &  =i\tau^{-\mbox{$\frac{1}{2}$}}-\frac{3i}{16}+\frac{155}{4608}\tau
^{\mbox{$\frac{1}{2}$}}+\ldots\nonumber\\
\phi &  =2\sqrt{\frac{2}{3}}i\tau^{\mbox{$\frac{1}{2}$}}-\frac{i}{16}%
\sqrt{\frac{2}{3}}\tau^{\mbox{$\frac{3}{2}$}}+\frac{31}{3204}\sqrt{\frac{2}%
{3}}\tau^{\mbox{$\frac{5}{2}$}}\ldots. \label{9.11}%
\end{align}

Neither pattern of dominant behaviour passes the Painlev\'e test
and so the system (\ref{9.7}) does not posses the Painlev\'e
Property.

\section{Scaling solutions}

Billyard \textit{et al }\ \cite{Billyard} discuss the stability of
cosmological scaling solutions within the class of spatially homogeneous
cosmological models with a perfect fluid subject to a linear equation of state
and a scalar field with an exponential potential. This type of study was
extended by van den Hoogen \textit{et al }\ \cite{Hoogen} to Robertson-Walker
spacetimes with nonzero curvature. Miritzis \textit{et al }\ \cite{Miritzis
00} have already shown that the corresponding problems in flat space possess
the Painlev\'{e} Property. For the former problem Billyard \textit{et al
}\ reduced the governing equations to the three-dimensional system%
\begin{align}
\dot{x} &  =\sqrt{\frac{3}{2}}\kappa y^{2}+\frac{3}{2}x\left[  \left(
\gamma-2\right)  \Omega-2y^{2}\right]  \nonumber\\
\dot{y} &  =3y-\sqrt{\frac{3}{2}}\kappa xy+\frac{3}{2}y\left[  \left(
\gamma-2\right)  \Omega-2y^{2}\right]  \nonumber\\
\dot{\Omega} &  =-3(\gamma-2)\Omega+3\Omega\left[  (\gamma-2)\Omega
-2y^{2}\right]  .\label{6.1}%
\end{align}

System (\ref{6.1}) has a superficial resemblance to a cubic system, but this
removed by the nonhomeographic transformation $y^{2}=z$ which reduces
(\ref{6.1}) to the quadratic system
\begin{align}
\dot{x}  &  =\sqrt{\frac{3}{2}}\kappa z+\frac{3}{2}x\left[  \left(
\gamma-2\right)  \Omega-2z\right] \nonumber\\
\dot{z}  &  =6z-\sqrt{6}\kappa xz+3z\left[  (\gamma-2)\Omega-2z\right]
\nonumber\\
\dot{\Omega}  &  =-3(\gamma-2)\Omega+3\Omega\left[  (\gamma-2)\Omega
-2z\right]  . \label{6.2}%
\end{align}
Since the transformation between the two systems is not homeographic, the
possession of the Painlev\'{e} Property by the one does not guarantee its
possession by the other. However, possession of it by (\ref{6.2}) will mean
that (\ref{6.1}) will possess the so-called weak Painlev\'{e} Property since
the transformation replaces a polelike singularity in $z$ with a branch point
singularity in $y$.

The standard substitution
\begin{equation}
x=\alpha\tau^{p},\quad z=\beta\tau^{q},\quad\Omega=\Gamma\tau^{r} \label{6.3}%
\end{equation}
into the derivatives and quadratic terms of (\ref{6.2}) gives the set of
exponents
\begin{equation}%
\begin{array}
[c]{cccc}%
p-1 & q & p+r & p+q\\
q-1 & p+q & q+r & 2q\\
r-1 & r & 2r & q+r.
\end{array}
\label{6.4}%
\end{equation}
The third of (\ref{6.4}) gives $q=r=-1$ and the second $p=-1$. The
coefficients of the leading order terms are found from the solution of the
linear system
\begin{align}
-\alpha &  =\frac{3}{2}(\gamma-2)\alpha\Gamma-3\alpha\beta\nonumber\\
-\beta &  =\sqrt{6}\kappa\alpha\beta+3(\gamma-2)\beta\Gamma-6\Gamma
^{2}\nonumber\\
-\Gamma &  =3(\gamma-2)\Gamma^{2}-6\beta\Gamma. \label{6.5}%
\end{align}
For nontrivial coefficients $\alpha$, $\beta$ and $\Gamma$ we see immediately
that there is an inconsistency between the first and third of (\ref{6.5}).
This can be resolved by setting either $p=0$ or $r=0$. However, the latter
choice leads to further inconsistencies and we are forced to the set $p=0$,
$q=r=-1$ and so the system (\ref{6.1}) cannot possess the Painlev\'{e} Property.

One could substitute the partially singular expansion
\begin{equation}
x=\sum_{i=0}a_{i}\tau^{i},\quad y=\sum_{i=0}b_{i}\tau^{(i-1)/2},\quad
\Omega=\sum_{i=0}c_{i}\tau^{i-1} \label{6.6}%
\end{equation}
into (\ref{6.1}). The results are somewhat disappointing. The coefficients in
the expansions can be expressed in terms of the one arbitrary coefficient,
$b_{0}$. In particular
\begin{equation}
a_{0}=\kappa b_{0}^{2}\quad\mbox{\rm and}\quad c_{0}=\frac{-1+b_{0}^{2}%
}{3(\gamma-2)}. \label{6.7}%
\end{equation}
The series for $y$ consists of even terms only. The solution depends upon two
arbitrary constants only, $b_{0}$ and $t_{0}$, and so is a peculiar solution.

The system derived for a self-interacting scalar field with an exponential
energy density evolving in a Robertson-Walker spacetime containing a
separately conserved perfect fluid is \cite{Billyard,Hoogen}
\begin{align}
\dot{x}  &  =-3x+\sqrt{\frac{3}{2}}\kappa y^{2}+\frac{1}{2}x\left[  \left(
3\gamma-2\right)  \Omega+2\left(  1+2x^{2}-y^{2}\right)  \right] \nonumber\\
\dot{y}  &  =-\sqrt{\frac{3}{2}}\kappa xy+\frac{1}{2}y\left[  \left(
3\gamma-2\right)  \Omega+2\left(  1+2x^{2}-y^{2}\right)  \right] \nonumber\\
\dot{\Omega}  &  =-3\gamma\Omega+\Omega\left[  \left(  3\gamma-2\right)
\Omega+2\left(  1+2x^{2}-y^{2}\right)  \right]  \label{6.8}%
\end{align}
in a standard notation. This system, with $-K$ in place of $\kappa$, has
already been given an extensive treatment in \cite{Miritzis 00} and we simply
summarise the results for completeness. The leading order exponents are
$p=q=\mbox{$\frac{1}{2}$}$ and $r=-1$. Under the transformation
\begin{equation}
u=4x^{2},\quad v=-2y^{2}\quad\mbox{\rm and}\quad w=(3\gamma-2)\Omega
\label{6.9}%
\end{equation}
the dominant terms of (\ref{6.8}) constitute the system
\begin{align}
\dot{u}  &  =u\left(  u+v+w\right) \nonumber\\
\dot{v}  &  =v\left(  u+v+w\right) \label{6.10}\\
\dot{w}  &  =w\left(  u+v+w\right) \nonumber
\end{align}
which has the leading order behaviour \cite{Miritzis 00}
\begin{equation}
u=u_{0}\tau^{-1},\quad v=v_{0}\tau^{-1},\quad w=w_{0}\tau^{-1}, \label{6.11}%
\end{equation}
where the coefficients are related according to $u_{0}+v_{0}+w_{0}+1=0$. That
the resonances are $r=-1,0(2)$ confirms that two of the coefficients of the
leading order terms are arbitrary and these with $t_{0}$ provide the three
arbitrary constants needed for the solution.

Returning to the original system (\ref{6.8}) we have that the system passes
the Painlev\'{e} test for the weak property. (There is no need to be concerned
about incompatibilities at the resonances since the arbitrary constants
already appear in the leading order terms. There are other patterns of
quasisingular behaviour similar to those listed in the table in \cite{Miritzis
00}. These do not fall within the ambit of the Painlev\'{e} test and do not
affect the possession by (\ref{6.8}) of the weak Painlev\'{e} Property and so
integrability in terms of functions with only branch point singularities (in
the case of $x$ and $y$; $\Omega$ is analytic away from its simple pole).

\section{Comments and conclusions \label{concl}}

The analysis of the systems of differential equations in the previous Sections
enables us to make some more general comments about the Painlev\'{e} property
and its possible applications. Firstly, we note that all the systems
considered here come from reductions of the Einstein equations of general
relativity by the imposition of certain symmetry and other plausible (from the
physical point of view) assumptions. That is, they correspond or represent
\emph{relativistic} systems. One major characteristic of such systems of
differential equations is their covariance, \emph{ie} the independence of
their properties from the coordinate frame used. However, it may be possible
that the use of the Painlev\'{e} analysis for a system written in different
coordinates leads to different behaviour. Thus we might conclude that the
Painlev\'{e} test is not an appropriate one to use when dealing with
relativistic equations. One may argue then that this test stands at a similar
level as that of using numerical methods conducive to the determination of the
so-called Liapunov exponents to decide on the issue of chaoticity of certain
generic classes of relativistic cosmologies.

Such an opinion is formulated, as the astute reader will appreciate, somewhat
tongue in cheek. The literature abounds with studies devoted to the
determination of closed form solutions of the Einstein equations for various
models. The variables of the equations are real and, not surprisingly, the
solutions sought are real. A real solution need not be analytic and yet be
quite useful for the explication of a specific physical model. Painlev\'{e}
requires solutions which are analytic in the complex plane of the independent
variable. This is a much stronger condition and, as we have remarked above, is
superficially in conflict with the concept of covariance in the field
equations and whatever is derived from them. The covariant formulation has no
regard for the preservation of analyticity. However, the Painlev\'{e} Property
is not preserved under a general transformation. Consequently what one should
be seeking amongst the plethora of possible coordinate representations is that
frame in which the Painlev\'{e} Property can be found, if it is ever to exist
for a given system. This, of course, raises a new line of research which it is
not appropriate to follow in this paper and that is to determine which of the
possible myriads of coordinate representations of the field equations of a
given model will be that set of coordinates for which the Painlev\'{e}
Property holds. Given that the system is integrable in terms of analytic
functions in one frame, the lesser demands of Cosmology can be met in a wider
variety of coordinate systems.

\section*{Acknowledgments}

PGLL thanks the Director of GEODYSYC, Dr S Cotsakis, and the Department of
Mathematics of the University of the Aegean for the provision of facilities
during the course of this work and acknowledges the continuing support of the
National Research Foundation of South Africa and the University of Natal.

\end{document}